\documentclass[twoside,slac_one]{revtex4}
\usepackage{graphicx}
\usepackage{fancyhdr}
\usepackage{amsmath} 
\usepackage{bm}
\usepackage{amsxtra}
\usepackage{amssymb}
\usepackage{amsthm}
\usepackage{latexsym}
\usepackage{lscape}

\begin{document}

\title{Nonstandard Higgs Decays and Dark Matter in the E$_6$SSM}

%

\author{J. P. Hall}
\author{S. F. King}
\affiliation{School of Physics and Astronomy, University of Southampton, Southampton, UK\\}

\author{R. Nevzorov}
\author{S. Pakvasa}
\affiliation{Department of Physics and Astronomy, University of Hawaii, Honolulu, USA\\}

\author{M. Sher}
\affiliation{Physics Department, College of William and Mary, Williamsburg, USA}

\begin{abstract}
We study the decays of the lightest Higgs boson within the exceptional
supersymmetric (SUSY) standard model (E$_6$SSM). The E$_6$SSM predicts 
three families of Higgs--like doublets plus three SM singlets that carry
$U(1)_{N}$ charges. One family of Higgs--like doublets and one SM singlet 
develop vacuum expectation values. The fermionic partners of other 
Higgs--like fields and SM singlets form inert neutralino and chargino 
states. Two lightest inert neutralinos tend to be the lightest and 
next-to-lightest SUSY particles (LSP and NLSP). The considered model 
can account for the dark matter relic abundance if the lightest inert 
neutralino has mass close to half the $Z$ mass. In this case the usual 
SM-like Higgs boson decays more than 95\% of the time into either LSPs 
or NLSPs. As a result the decays of the lightest Higgs boson into
$l^{+} l^{-} + X$ might play an essential role in the Higgs searches. 
This scenario also predicts other light inert chargino and neutralino 
states below $200\,\mbox{GeV}$ and large LSP direct detection 
cross-sections which is on the edge of observability of XENON100.
\end{abstract}

\maketitle

\thispagestyle{fancy}


\section{Introduction}
Confirming the Higgs mechanism as the underlying principle of electroweak 
symmetry breaking is one of the main goals of upcoming accelerators.
The strategy for Higgs boson searches depends on the production
mechanism and on the decay branching fractions of Higgs to 
different channels. Physics beyond the Standard Model may
lead to the modification of the Higgs signals. In particular, there exist 
several extensions of the Standard Model in which the Higgs boson can decay 
with a substantial branching fraction into particles which can not be 
directly detected. The presence of invisible decays modifies considerably
Higgs boson searches, making Higgs discovery much more difficult.
If the Higgs is mainly invisible, then the visible branching ratios will 
be dramatically reduced, preventing detection in the much studied channels 
at the LHC and the Tevatron. At $e^{+} e^{-}$ colliders, the problems related 
to the observation of the invisible Higgs are less severe since it can be 
tagged through the recoiling $Z$. As a result the LEP II collaborations 
exclude invisible Higgs masses up to 114.4 GeV \cite{:2001xz}. On the other 
hand, Higgs searches at hadron colliders are more difficult in the presence 
of such invisible decays. Previous studies have analysed $ZH$ and $WH$ 
associated production \cite{Choudhury:1993hv}-\cite{Davoudiasl:2004aj} as 
well as $t\bar{t} H$ production \cite{Gunion:1993jf}-\cite{2} and $t\bar{t} VV$ 
($b\bar{b} VV$) production \cite{Boos:2010pu} as promising channels.

Here we consider the exotic decays of the lightest Higgs boson and associated 
novel collider signatures within the Exceptional Supersymmetric Standard Model (E$_6$SSM).
The E$_6$SSM is based on the $SU(3)_C\times SU(2)_W\times U(1)_Y \times U(1)_N$
gauge group which is a subgroup of $E_6$ \cite{King:2005jy}-\cite{King:2005my}. 
The additional low energy $U(1)_N$ is a linear superposition of $U(1)_{\chi}$ 
and $U(1)_{\psi}$, i.e. 
\begin{equation}
U(1)_N=\frac{1}{4} U(1)_{\chi}+\frac{\sqrt{15}}{4} U(1)_{\psi}\,.
\label{essm1}
\end{equation}
The two anomaly--free $U(1)_{\psi}$ and $U(1)_{\chi}$ symmetries are defined by:
$E_6\to SO(10)\times U(1)_{\psi}\,,\quad SO(10)\to SU(5)\times U(1)_{\chi}$.
The extra $U(1)_N$ gauge symmetry is defined such that right--handed
neutrinos do not participate in the gauge interactions. Since right--handed 
neutrinos have zero charges they can be superheavy, shedding light on the 
origin of the mass hierarchy in the lepton sector and providing a mechanism 
for the generation of the baryon asymmetry in the Universe via 
leptogenesis \cite{King:2008qb}.

To ensure anomaly cancellation the particle content of the E$_6$SSM is 
extended to include three complete fundamental $27$ representations of $E_6$. 
Each $27_i$ multiplet contains a SM family of quarks and leptons, 
right--handed neutrino $N^c_i$, SM-type singlet fields $S_i$ which carry 
non-zero $U(1)_{N}$ charge, a pair of $SU(2)_W$--doublets $H^d_{i}$ and 
$H^u_{i}$ and a pair of colour triplets of exotic quarks $\overline{D}_i$ and 
$D_i$ which can be either diquarks (Model I) or leptoquarks (Model II)
\cite{King:2005jy}-\cite{King:2005my}.\, The $S_i$, $H^d_{i}$ and $H^u_{i}$ 
form either Higgs or inert Higgs multiplets. In addition to the complete $27_i$ 
multiplets the low energy particle spectrum of the E$_6$SSM is supplemented 
by $SU(2)_W$ doublet $L_4$ and anti-doublet $\overline{L}_4$ states from 
extra $27'$ and $\overline{27'}$ to preserve gauge coupling unification.
The analysis performed in \cite{King:2007uj} shows 
that the unification of gauge couplings in the E$_6$SSM can be achieved 
for any phenomenologically acceptable value of $\alpha_3(M_Z)$ consistent 
with the measured low energy central value. The presence of a $Z'$ boson 
and of exotic quarks predicted by the E$_6$SSM provides spectacular new 
physics signals at the LHC which were discussed in \cite{King:2005jy}--\cite{King:2005my}, 
\cite{King:2006vu}--\cite{King:2006rh}. Recently the particle 
spectrum and collider signatures associated with it were studied within 
the constrained version of the E$_6$SSM \cite{Athron:2008np}--\cite{Athron:2011wu}.

The superpotential in the $E_6$ inspired models involves many new Yukawa 
couplings that induce non--diagonal flavour transitions. To suppress these effects 
in the E$_6$SSM an approximate $Z^{H}_2$ symmetry is imposed. Under this symmetry
all superfields except one pair of $H^d_{i}$ and $H^u_{i}$ (say $H_d\equiv H^d_{3}$ and
$H_u\equiv H^u_{3}$) and one SM-type singlet field ($S\equiv S_3$) are odd. The $Z^{H}_2$ 
symmetry reduces the structure of the Yukawa interactions to
\begin{eqnarray}
W_{\rm E_6SSM}&\simeq &  \lambda \hat{S} (\hat{H}_u \hat{H}_d)+
\lambda_{\alpha\beta} \hat{S} (\hat{H}^d_{\alpha} \hat{H}^u_{\beta})
+\tilde{f}_{\alpha\beta} \hat{S}_{\alpha} (\hat{H}^d_{\beta}\hat{H}_u)
+f_{\alpha\beta} \hat{S}_{\alpha} (\hat{H}_d \hat{H}^u_{\beta})
+\kappa_{ij} \hat{S} (\hat{D}_i\hat{\overline{D}}_j)
\nonumber\\[2mm]
&+&
h^U_{ij}(\hat{H}_{u} \hat{Q}_i)\hat{u}^c_{j} + h^D_{ij}(\hat{H}_{d} \hat{Q}_i)\hat{d}^c_j
+ h^E_{ij}(\hat{H}_{d} \hat{L}_i)\hat{e}^c_{j}+ h_{ij}^N(\hat{H}_{u} \hat{L}_i)\hat{N}_j^c\nonumber\\[2mm]
&+& \dfrac{1}{2}M_{ij}\hat{N}^c_i\hat{N}^c_j+\mu'(\hat{L}_4\hat{\overline{L}}_4)+
h^{E}_{4j}(\hat{H}_d \hat{L}_4)\hat{e}^c_j
+h_{4j}^N(\hat{H}_{u}\hat{L}_4)\hat{N}_j^c\,,
\label{essm2}
\end{eqnarray}
where $\alpha,\beta=1,2$ and $i,j=1,2,3$\,. 
The $SU(2)_W$ doublets $\hat{H}_u$ and $\hat{H}_d$
and SM-type singlet field $\hat{S}$, that are even under the $Z^{H}_2$ symmetry,
play the role of Higgs fields. At the physical vacuum they develop vacuum expectation 
values (VEVs)
\begin{equation}
\langle H_d\rangle =\displaystyle\frac{1}{\sqrt{2}}\left(
\begin{array}{c}
v_1\\ 0
\end{array}
\right) , \qquad
\langle H_u\rangle =\displaystyle\frac{1}{\sqrt{2}}\left(
\begin{array}{c}
0\\ v_2
\end{array}
\right) ,\qquad
\langle S\rangle =\displaystyle\frac{s}{\sqrt{2}}.
\label{essm3}
\end{equation}
generating the masses of the quarks and leptons. Instead of $v_1$ and $v_2$ it is more 
convenient to use $\tan\beta=v_2/v_1$ and $v=\sqrt{v_1^2+v_2^2}=246\,\mbox{GeV}$. 
The VEV of the SM-type singlet field, $s$, breaks the extra $U(1)_N$ symmetry 
thereby providing an effective $\mu$ term as well as the necessary exotic fermion 
masses and also inducing that of the $Z'$ boson. In the E$_6$SSM the Higgs 
spectrum contains one pseudoscalar, two charged and three CP--even states.
In the leading two--loop approximation the mass of the lightest CP--even 
Higgs boson does not exceed $150-155\,\mbox{GeV}$ \cite{King:2005jy}.

Although $Z^{H}_2$ eliminates any problems related with baryon number violation 
and non-diagonal flavour transitions it also forbids all Yukawa interactions that 
would allow the exotic quarks to decay. Since models with stable charged exotic 
particles are ruled out by various experiments the $Z^{H}_2$ symmetry can only 
be an approximate one. From here on we assume that $Z^{H}_2$ symmetry violating 
couplings are small and can be neglected in our analysis. This assumption can be 
justified if we take into account that the $Z^{H}_2$ symmetry violating operators 
may give an appreciable contribution to the amplitude of $K^0-\overline{K}^0$ 
oscillations and give rise to new muon decay channels like $\mu\to e^{-}e^{+}e^{-}$. 
In order to suppress processes with non--diagonal flavour transitions the Yukawa 
couplings of the exotic particles to the quarks and leptons of the first two 
generations should be smaller than $10^{-3}-10^{-4}$. Such small $Z^{H}_2$ symmetry 
violating couplings can be ignored in the first approximation.


\section{Masses and couplings of the lightest inert neutralinos}
When $Z^{H}_2$ symmetry violating couplings tend to zero only $H_u$, $H_d$
and $S$ acquire non-zero VEVs. In this approximation the charged components 
of the inert Higgsinos 
$(\tilde{H}^{u+}_2,\,\tilde{H}^{u+}_1,\,\tilde{H}^{d-}_2,\,\tilde{H}^{d-}_1)$
and ordinary chargino states do not mix. The neutral components of the inert 
Higgsinos
($\tilde{H}^{d0}_1$, $\tilde{H}^{d0}_2$, $\tilde{H}^{u0}_1$, $\tilde{H}^{u0}_2$)
and inert singlinos ($\tilde{S}_1$, $\tilde{S}_2$) also do not mix with
the ordinary neutralino states. Moreover if $Z^{H}_2$ symmetry was exact then 
both the lightest state in the ordinary neutralino sector and the lightest 
inert neutralino would be absolutely stable. Therefore, although $Z^{H}_2$ 
symmetry violating couplings are expected to be rather small, we shall
assume that they are large enough to allow either the lightest neutralino 
state or the lightest inert neutralino to decay within a reasonable time.

In the field basis
$(\tilde{H}^{d0}_2,\,\tilde{H}^{u0}_2,\,\tilde{S}_2,\,\tilde{H}^{d0}_1,\,\tilde{H}^{u0}_1,\,\tilde{S}_1)$
the mass matrix of the inert neutralino sector takes a form
\begin{equation}
M_{IN}=
\left(
\begin{array}{cc}
A_{22}  & A_{21} \\[2mm]
A_{12}  & A_{11}
\end{array}
\right)\,,\qquad 
A_{\alpha\beta}=-\frac{1}{\sqrt{2}}
\left(
\begin{array}{ccc}
0                                           & \lambda_{\alpha\beta} s           & \tilde{f}_{\beta\alpha} v \sin{\beta}
\\[2mm]
\lambda_{\beta\alpha} s                     & 0                                 & f_{\beta\alpha} v \cos{\beta} \\[2mm]
\tilde{f}_{\alpha\beta} v \sin{\beta}       & f_{\alpha\beta} v \cos{\beta}     & 0
\end{array}
\right)\,,  
\label{icn1}
\end{equation}
where $A_{\alpha\beta}$ are $3\times 3$ sub-matrices and $A_{12}=A^{T}_{21}$.
In the basis of inert chargino interaction states
$(\tilde{H}^{u+}_2,\,\tilde{H}^{u+}_1,\,\tilde{H}^{d-}_2,\,\tilde{H}^{d-}_1)$
the corresponding mass matrix can be written as
\begin{equation}
M_{IC}=
\left(
\begin{array}{cc}
0  & C^{T} \\[2mm]
C  & 0
\end{array}
\right)\,,\qquad C_{\alpha\beta}=\dfrac{1}{\sqrt{2}}\lambda_{\alpha\beta}\, s\,,
\label{icn3}
\end{equation}
where $C_{\alpha\beta}$ are $2\times 2$ sub-matrices. 
 
In our analysis we require the validity of perturbation theory up to the 
GUT scale that constrains the allowed range of all Yukawa couplings.
We also choose $s$ and $\lambda_{\alpha\beta}$ so that the masses of all 
inert chargino states are larger than $100\,\mbox{GeV}$ and $Z'$ boson
is relatively heavy ($M_{Z'}\gtrsim 865\,\mbox{GeV}$). The restrictions 
specified above set very strong limits on the masses of the lightest 
inert neutralinos. In particular, our numerical analysis indicates that 
the lightest and second lightest inert neutralinos ($\chi^0_{1}$ and $\chi^0_{2}$) 
are typically lighter than $60-65\,\mbox{GeV}$ \cite{10}-\cite{Hall:2010ny}. 
Therefore the lightest inert neutralino tends to be the lightest SUSY particle 
in the spectrum and can play the role of dark matter. The neutralinos 
$\chi^0_{1}$ and $\chi^0_{2}$ are predominantly inert singlinos. Their 
couplings to the $Z$--boson can be rather small so that such inert neutralinos 
would remain undetected at LEP. 

In order to clarify the results of our numerical analysis, it is useful to 
consider a simple scenario when
$$
\lambda_{\alpha\beta}=\lambda_{\alpha}\,\delta_{\alpha\beta},\qquad
f_{\alpha\beta}=f_{\alpha}\,\delta_{\alpha\beta},\qquad
\tilde{f}_{\alpha\beta}=\tilde{f}_{\alpha}\,\delta_{\alpha\beta}\,.
$$
In this case the mass matrix of inert neutralinos reduces to the block 
diagonal form. In the limit where $f_{\alpha}=\tilde{f}_{\alpha}$ one can easily 
prove using the method proposed in \cite{Hesselbach:2007te} that there are theoretical 
upper bounds on the masses of the lightest and second lightest inert neutralino states. 
The corresponding theoretical restrictions are
\begin{equation}
|m_{\chi^0_{\alpha}}|^2\lesssim \mu_{\alpha}^2=\dfrac{1}{2}\biggl[|m_{\chi^{\pm}_{\alpha}}|^2+
\dfrac{f_{\alpha}^2 v^2}{2}
\biggl(1+\sin^22\beta\biggr)-
\sqrt{\biggl(|m_{\chi^{\pm}_{\alpha}}|^2+
\dfrac{f_{\alpha}^2 v^2}{2}(1+\sin^22\beta)\biggr)^2-
f_{\alpha}^4 v^4 \sin^22\beta}\biggr]\,,
\label{icn6}
\end{equation}
where $m_{\chi^{\pm}_{\alpha}}=\lambda_{\alpha} s/\sqrt{2}$ are the masses of the inert charginos.
The value of $\mu_{\alpha}$ decreases with increasing $|m_{\chi^{\pm}_{\alpha}}|$ and $\tan\beta$. 
At large values of $|m_{\chi^{\pm}_{\alpha}}|$ and $\tan\beta$, Eq.~(\ref{icn6}) 
simplifies resulting in
\begin{equation}
|m_{\chi^0_{\alpha}}|^2\lesssim \dfrac{f_{\alpha}^4 v^4 \sin^22\beta}{4\biggl(|m_{\chi^{\pm}_{\alpha}}|^2+
\dfrac{f_{\alpha}^2 v^2}{2}(1+\sin^22\beta)\biggr)}\,.
\label{icn7}
\end{equation}
The theoretical restriction on $|m_{\chi^0_{\alpha}}|$ achieves its maximal value around 
$\tan\beta\simeq 1.5$. For this value of $\tan\beta$ the requirement of the validity of 
perturbation theory up to the GUT scale implies that $f_{1}=\tilde{f}_{1}=f_{2}=\tilde{f}_{2}$ 
are less than $0.6$. As a consequence the lightest inert neutralinos are
lighter than $60-65\,\mbox{GeV}$ for $|m_{\chi^{\pm}_{\alpha}}|> 100\,\mbox{GeV}$.

The inert neutralino mass matrix (\ref{icn1}) can be diagonalized using the neutralino mixing
matrix defined by
\begin{equation}
N_i^a M^{ab} N_j^b = m_i \delta_{ij}, \qquad\mbox{ no sum on } i.
\label{icn8}
\end{equation}
In the limit where off--diagonal Yukawa couplings vanish and 
$\lambda_{\alpha} s\gg f_{\alpha} v,\, \tilde{f}_{\alpha} v$
the eigenvalues of the inert neutralino mass matrix can be easily 
calculated (see \cite{Hall:2009aj}). In particular, the masses of two 
lightest inert neutralino states ($\chi^0_{1}$ and $\chi^0_{2}$)
are given by
\begin{equation}
m_{\chi^0_{\alpha}}\simeq \frac{\tilde{f}_{\alpha} f_{\alpha} v^2 \sin 2\beta}{2\, m_{\chi^{\pm}_{\alpha}}}\,.
\label{icn10}
\end{equation}
From Eq.~(\ref{icn10}) one can see that the masses of $\chi^0_{1}$ and $\chi^0_{2}$ 
are determined by the values of the Yukawa couplings $\tilde{f}_{\alpha}$ and $f_{\alpha}$. 

The lightest inert neutralino states are made up of the following superposition 
of interaction states
\begin{equation}
\tilde{\chi}_{\alpha}^0 = N_{\alpha}^1 \tilde{H}^{d0}_2 + N_{\alpha}^2 \tilde{H}^{u0}_2 + N_{\alpha}^3 \tilde{S}_2 +
N_{\alpha}^4 \tilde{H}^{d0}_1 + N_{\alpha}^5 \tilde{H}^{u0}_1 + N_{\alpha}^6 \tilde{S}_1\,,
\label{icn11}
\end{equation}
Using the above lightest and second lightest inert neutralino compositions
it is straightforward to derive the couplings of these states to the $Z$-boson.
In general the part of the Lagrangian that describes the interactions of $Z$ 
with $\chi^0_1$ and $\chi^0_2$, can be presented in the following form:
\begin{equation}
\mathcal{L}_{Z\chi\chi}=\sum_{\alpha,\beta}\dfrac{M_Z}{2 v}Z_{\mu}
\biggl(\overline{\chi^{0}_{\alpha}}\gamma_{\mu}\gamma_{5}\chi^0_{\beta}\biggr) R_{Z\alpha\beta}\,,\qquad\qquad
R_{Z\alpha\beta}=N_{\alpha}^1 N_{\beta}^1 - N_{\alpha}^2 N_{\beta}^2 + N_{\alpha}^4 N_{\beta}^4 -
N_{\alpha}^5 N_{\beta}^5\,.
\label{icn13}
\end{equation}
In the case where off--diagonal Yukawa couplings go to zero while
$\lambda_{\alpha} s\gg f_{\alpha} v,\, \tilde{f}_{\alpha} v$
the relative couplings of the lightest and second lightest inert
neutralino states to the $Z$-boson are given by
\begin{equation}
R_{Z\alpha\beta}=R_{Z\alpha\alpha}\,\delta_{\alpha\beta}\,,\qquad
R_{Z\alpha\alpha}=\dfrac{v^2}{2 m_{\chi^{\pm}_{\alpha}}^2}
\biggl(f_{\alpha}^2\cos^2\beta-\tilde{f}_{\alpha}^2\sin^2\beta\biggr)\,.
\label{icn14}
\end{equation}
One can see that the couplings of $\chi^0_1$ and $\chi^0_2$ to the 
$Z$-boson can be very strongly suppressed or even tend to zero. This 
happens when $|f_{\alpha}|\cos\beta=|\tilde{f}_{\alpha}|\sin\beta$,
which is when $\chi^0_{\alpha}$ contains a completely symmetric 
combination of $\tilde{H}^{d0}_{\alpha}$ and $\tilde{H}^{u0}_{\alpha}$. 
Eq.~(\ref{icn14}) also indicates that the couplings of $\chi^0_1$ and 
$\chi^0_2$ to $Z$ are always small when inert charginos are rather 
heavy or $\tilde{f}_{\alpha}$ and $f_{\alpha}$ are small
(i.e. $m_{\chi^0_{\alpha}}\to 0$).

Although $\chi^0_1$ and $\chi^0_2$ might have extremely small couplings to 
$Z$, their couplings to the lightest CP--even Higgs boson $h_1$ can not be 
negligibly small if the corresponding states have appreciable masses. 
If all Higgs states except the lightest one are considerably heavier 
than the EW scale the mass matrix of the CP--even Higgs sector can be 
diagonalised using the perturbation theory 
\cite{Nevzorov:2001um}-\cite{Nevzorov:2004ge}. Then the effective Lagrangian 
that describes the interactions of the inert neutralinos with the lightest 
CP-even Higgs eigenstate takes the form
\begin{equation}
\mathcal{L}_{h_1\chi\chi}\simeq\sum_{i,j} (-1)^{\theta_i+\theta_j} X^{h_1}_{ij} \biggl(\psi^{0T}_{i}
(-i\gamma_{5})^{\theta_i+\theta_j}\psi^0_{j}\biggr) h_1\,,\qquad\qquad
X^{h_1}_{ij}=-\dfrac{1}{\sqrt{2}}\biggl(F_{ij}\cos\beta+\tilde{F}_{ij}\sin\beta\biggr)\,,
\label{higgs7}   
\end{equation}
where $i,j=1,2,...6$ and
$$
\begin{array}{c}
F_{ij} = f_{11} N^6_i N^5_j + f_{12} N^6_i N^2_j + f_{21} N^3_i N^5_j +f_{22} N^3_i N^2_j\,,\\[2mm] 
\tilde{F}_{ij} = \tilde{f}_{11} N^6_i N^4_j + \tilde{f}_{12} N^6_i N^1_j + 
\tilde{f}_{21} N^3_i N^4_j + \tilde{f}_{22} N^3_i N^1_j\,.
\end{array}
$$
In Eq.~(\ref{higgs7}) $\psi^0_i=(-i\gamma_5)^{\theta_i}\chi^0_i$ is
the set of inert neutralino eigenstates with positive eigenvalues, while $\theta_i$ equals 
0 (1) if the eigenvalue corresponding to $\chi^0_i$ is positive (negative). The inert 
neutralinos are labeled according to increasing absolute value of mass, with $\psi^0_1$ 
being the lightest inert neutralino and $\psi^0_6$ the heaviest.

In the limit when off-diagonal Yukawa couplings that determine the interactions
of the inert Higgs fields with $H_u$, $H_d$ and $S$ vanish and 
$\lambda_{\alpha} s\gg f_{\alpha} v,\, \tilde{f}_{\alpha} v$) one obtains
\begin{equation}
X^{h_1}_{\alpha\beta}\simeq\dfrac{|m_{\chi^0_{\alpha}}|}{v}\,\delta_{\alpha\beta}\,,
\label{higgs9}
\end{equation}
where $\alpha,\beta=1,2$, labeling the two light, mostly inert singlino states.
These simple analytical expressions for the couplings of the SM--like Higgs boson to 
the lightest and second lightest inert neutralinos are not as surprising as they may 
first appear. When the Higgs spectrum is hierarchical, the VEV of the lightest CP--even 
state is responsible for all light fermion masses in the E$_6$SSM. As a result we 
expect that their couplings to SM--like Higgs can be written as usual as being proportional 
to the mass divided by the VEV.

\section{Exotic Higgs decays and Dark Matter}
In our analysis we require that the lightest inert neutralino account for all 
or some of the observed dark matter relic density, which is measured to be 
$\Omega_{\mathrm{CDM}}h^2 = 0.1099 \pm 0.0062$ \cite{cdm}. If a theory predicts
a greater relic density of dark matter than this then it is ruled out, assuming 
standard pre-BBN cosmology. A theory that predicts less dark matter cannot be 
ruled out in the same way but then there would have to be other contributions 
to the dark matter relic density.

In the limit where all non-SM fields other than the two lightest inert neutralinos
are heavy ($\gtrsim \mbox{TeV}$) the lightest inert neutralino state in the 
E$_6$SSM results in too large density of dark matter. Indeed, because
the mass of $\tilde{\chi}^0_1$ is inversely proportional to the masses of inert 
charginos the lightest inert neutralinos tend to be very light 
$|m_{\chi^0_{\sigma}}|\ll M_Z$. As a result the couplings of $\tilde{\chi}^0_1$
to gauge bosons, Higgs states, quarks (squarks) and leptons (sleptons) are quite 
small leading to a relatively small annihilation cross section for
$\tilde{\chi}^0_1\tilde{\chi}^0_1\to \mbox{SM particles}$. Since the dark matter 
number density is inversely proportional to the annihilation cross section at the 
freeze-out temperature the lightest inert neutralino state gives rise to a relic 
density that is typically much larger than its measured value. Thus in the limit 
considered the bulk of the E$_6$SSM parameter space that leads to small masses 
of $\tilde{\chi}^0_1$ is almost ruled out\footnote{When $f_{\alpha\beta},\, \tilde{f}_{\alpha\beta}\to 0$
the masses of $\tilde{\chi}^0_1$ and $\tilde{\chi}^0_2$ tend to zero and 
inert singlino states essentially decouple from the rest of the spectrum.
In this limit the lightest non-decoupled neutralino may be rather stable
and can play the role of dark matter \cite{Hall:2011zq}. The presence of very 
light neutral fermions in the particle spectrum might have interesting implications 
for the neutrino physics (see, for example \cite{Frere:1996gb}).}.

A reasonable density of dark matter can be obtained for 
$|m_{\chi^0_{1}}|\sim M_Z/2$ when the lightest inert neutralino states annihilate 
mainly through an $s$--channel $Z$--boson, via its inert Higgsino doublet components 
which couple to the $Z$--boson. It is worth noting that if $\tilde{\chi}^0_1$ was 
pure inert Higgsino then the $s$--channel $Z$--boson annihilation would proceed
with the full gauge coupling strength leaving the relic density too low to account 
for the observed dark matter. In the E$_6$SSM the LSP is mostly inert singlino so 
that its coupling to the $Z$--boson is typically suppressed, since it only couples 
through its inert Higgsino admixture leading to an increased relic density. In practice, 
the appropriate value of $\Omega_{\mathrm{CDM}}h^2$ can be achieved even if the coupling 
of $\tilde{\chi}^0_1$ to the $Z$--boson is relatively small. This happens when 
$\tilde{\chi}^0_1$ annihilation proceeds through the $Z$--boson resonance, i.e. 
$2|m_{\chi^0_{1}}|\simeq M_Z$. 

Because the scenarios that result in the reasonable density of dark matter imply that
$\chi^0_1$ and $\chi^0_2$ have large couplings to the lightest Higgs boson which are 
much larger than the $b$--quark Yukawa coupling and the decays of the lightest Higgs 
state into these inert neutralinos are kinematically allowed, the
SM--like Higgs boson decays predominantly into $\chi^0_1$ and $\chi^0_2$.
The corresponding partial decay widths are given by
\begin{equation}
\begin{array}{rcl}
\Gamma(h_1\to\chi^0_{\alpha}\chi^0_{\beta})&=&\dfrac{\Delta_{\alpha\beta}}{8\pi m_{h_1}}
\biggl(X^{h_1}_{\alpha\beta}+X^{h_1}_{\beta\alpha}\biggr)^2\biggl[
m^2_{h_1}-(|m_{\chi^0_{\alpha}}|+(-1)^{\theta_{\alpha}+\theta_{\beta}}|m_{\chi^0_{\beta}}|)^2\biggr]\\[4mm]
&&\times\sqrt{\biggl(1-\dfrac{|m_{\chi^0_{\alpha}}|^2}{m^2_{h_1}}-\dfrac{|m_{\chi^0_{\beta}}|^2}{m^2_{h_1}}\biggr)^2-
4\dfrac{|m_{\chi^0_{\alpha}}|^2 |m_{\chi^0_{\beta}}|^2}{m^4_{h_1}}}\,,
\end{array}
\label{higgs10}
\end{equation}
where $\Delta_{\alpha\beta}=\dfrac{1}{2}\,(1)$ for $\alpha=\beta$ ($\alpha\neq\beta$).
On the other hand the large coupling of $\chi^0_1$ to the lightest Higgs state give
rise to the relatively large $\chi_1^0$--nucleon elastic--scattering cross section.
Since in the E$_6$SSM the couplings of the lightest inert neutralino to quarks (leptons) 
and squarks (sleptons) are suppressed, the $\chi_1^0$--nucleon elastic scattering, 
which is associated with the spin-independent cross section, is mediated mainly by 
the $t$--channel lightest Higgs boson exchange. Thus in the leading approximation 
the spin--independent part of $\chi_1^0$--nucleon cross section in the
E$_6$SSM takes the form 
\begin{equation}
\begin{array}{l}
\sigma_{SI}=\dfrac{4 m^2_r m_N^2}{\pi v^2 m^4_{h_1}} |X^{h_1}_{11} F^N|^2\,,\\[4mm]
m_r=\dfrac{m_{\chi^0_1} m_N}{m_{\chi^0_1}+m_N}\,,\qquad\qquad\qquad
F^N=\sum_{q=u,d,s} f^N_{Tq} + \dfrac{2}{27}\sum_{Q=c,b,t} f^N_{TQ}\,,
\end{array}
\label{higgs17}
\end{equation}
where
$$
m_N f^N_{Tq} = \langle N | m_{q}\bar{q}q |N \rangle\,, \qquad\qquad\qquad\qquad
f^N_{TQ} = 1 - \sum_{q=u,d,s} f^N_{Tq}\,.
$$
As one can see from Eq.~(\ref{higgs17}) the value of $\sigma_{SI}$ depends
rather strongly on the hadronic matrix elements, i.e. the coefficients $f^N_{Tq}$, that
are related to the $\pi$--nucleon $\sigma$ term and the spin content of the nucleon.
As a consequence $\sigma_{SI}$ varies over a wide range (see Table 1).
Recently the CDMSII and XENON100 collaborations set upper limits on 
$\sigma_{SI}$ \cite{Ahmed:2009zw},\cite{Aprile:2010um}.

\begin{table}[ht]
\begin{center}
\caption{Benchmark scenarios. The branching ratios and decay widths of the
lightest Higgs boson, the masses of the Higgs states, inert neutralinos and 
charginos as well as the couplings of $\tilde{\chi}^0_1$ and $\tilde{\chi}^0_2$ 
are calculated for $s=2400\,\mbox{GeV}$, $m_Q=m_U=M_S=700\,\mbox{GeV}$, 
$X_t=\sqrt{6} M_S$ that correspond to $m_{h_2}\simeq M_{Z'}\simeq 890\,\mbox{GeV}$.}
\begin{tabular}{|c|c|c|c|c|c|}
\hline
                           &	i	    &	ii	 &	iii	    &  iv               & v\\
\hline
$\lambda$	           &	0.6	    &	0.6	 &   0.468       &	0.468	&	0.468	\\
\hline
$\tan(\beta)$	           &	1.7	    &	1.564    &	1.5	    & 1.5       & 1.5\\
\hline
$A_{\lambda}$	               &	1600	    &	1600 &   600    &	600	&	600	\\
\hline
$m_{H^{\pm}}\simeq m_{A}\simeq m_{h_3}$/GeV&1977    &	1990&  1145	& 1145                & 1145\\
\hline
$m_{h_1}$/GeV	           &	133.1	&	134.8	 &	115.9	& 115.9               & 115.9\\
\hline
$\lambda_{22}$	           &	0.094	&	0.0001	 &	0.094	& 0.001               & 0.468\\
\hline
$\lambda_{21}$	           &	0	    &	0.06	 &	0	    & 0.079               & 0.05\\
\hline
$\lambda_{12}$	           &	0	    &	0.06	 &	0	    & 0.080               & 0.05\\
\hline
$\lambda_{11}$	           &	0.059	&	0.0001	 &	0.059	& 0.001               & 0.08\\
\hline
$f_{22}$	           &	0.53	&	0.001	 &	0.53	& 0.04                & 0.05\\
\hline
$f_{21}$	           &	0.05	&	0.476	 &	0.053	& 0.68                & 0.9\\
\hline
$f_{12}$	           &	0.05	&	0.466	 &	0.053	& 0.68                & 0.002\\
\hline
$f_{11}$	           &	0.53	&	0.001	 &	0.53	& 0.04                & 0.002\\
\hline
$\tilde{f}_{22}$	   &	0.53	&	0.001	 &	0.53	& 0.04                & 0.002\\
\hline
$\tilde{f}_{21}$	   &	0.05	&	0.4	 &	0.053	& 0.49                & 0.002\\
\hline
$\tilde{f}_{12}$	   &	0.05	&	0.408	 &	0.053	& 0.49                & 0.05\\
\hline
$\tilde{f}_{11}$	   &	0.53	&	0.001	 &	0.53	& 0.04                & 0.65\\
\hline
$m_{\tilde{\chi}^0_1}$/GeV &	33.62	&	-36.69   &	35.42	& -45.08              & -46.24\\
\hline
$m_{\tilde{\chi}^0_2}$/GeV &	47.78	&	36.88	 &	51.77	& 55.34               & 46.60\\
\hline
$m_{\tilde{\chi}^0_3}$/GeV &	108.0	&	-103.11  &	105.3	& -133.3              & 171.1\\
\hline
$m_{\tilde{\chi}^0_4}$/GeV &	-152.1	&	103.47   &	-152.7	& 136.9               & -171.4\\
\hline
$m_{\tilde{\chi}^0_5}$/GeV &	163.5	&	139.80   &	162.0	& 178.4               & 805.4\\
\hline
$m_{\tilde{\chi}^0_6}$/GeV &	-200.8	&	-140.35  &	-201.7	& -192.2              & -805.4\\
\hline
$m_{\tilde{\chi}^\pm_1}$/GeV&	100.1	&	101.65	 &	100.1	& 133.0               & 125.0\\
\hline
$m_{\tilde{\chi}^\pm_2}$/GeV&	159.5	&	101.99	 &	159.5	& 136.8               & 805.0\\
\hline
$\Omega_\chi h^2$	    &	0.109	&	0.107	 &	0.107	& 0.0324              & 0.00005\\
\hline
$R_{Z11}$	            &	-0.144	&	-0.132	 &	-0.115	& -0.0217             & -0.0224\\
\hline
$R_{Z12}$	            &	0.051	&	0.0043	 &	-0.045	& -0.0020             & -0.213\\
\hline
$R_{Z22}$	            &	-0.331	&	-0.133	 &	-0.288	& -0.0524             & -0.0226\\
\hline
$\sigma_{SI}/10^{-44}$ cm$^2$&1.7-7.1        &2.0-8.2    &3.5-14.2       & 6.0-24.4                    & 6.1-25.0\\
\hline
$\mathrm{Br}(h\rightarrow \tilde{\chi}^0_1 \tilde{\chi}^0_1)$& 57.8\% & 49.1\%&76.3\%&83.4\%             &49.3\%\\
\hline
$\mathrm{Br}(h\rightarrow \tilde{\chi}^0_1 \tilde{\chi}^0_2)$& 0.34\% & $3.5\times 10^{-11}$
&0.26\% &$7.6\times 10^{-9}$ &$3.0\times 10^{-8}$\\
\hline
$\mathrm{Br}(h\rightarrow \tilde{\chi}^0_2 \tilde{\chi}^0_2)$& 39.8\% &49.2\% &20.3\%
&12.3\%   &47.9\%\\
\hline
$\mathrm{Br}(h\rightarrow b\bar{b})$                        & 1.87\% & 1.59\% &2.83\%&3.95\%             &2.58\%\\
\hline
$\mathrm{Br}(h\rightarrow \tau\bar{\tau})$                  & 0.196\%& 0.166\% &0.30\%&0.41\%             &0.27\%\\
\hline
$\Gamma^{tot}$/MeV                                          & 141.2  & 169.0 &82.0  &58.8               &90.1\\
\hline
\end{tabular}
\label{table-1}
\end{center}
\end{table}

In order to illustrate the features of the E$_6$SSM mentioned above we specify a set of 
benchmark points (see Table 1). For each benchmark scenario we calculate the spectrum of 
the inert neutralinos, inert charginos and Higgs bosons as well as the branching ratios 
of the decays of the lightest CP-even Higgs state and the dark matter relic density.
In Table 1 the masses of the heavy Higgs states are computed in the leading one--loop 
approximation. In the case of the lightest Higgs boson mass the leading two--loop 
corrections are taken into account. In order to construct benchmark scenarios that are 
consistent with cosmological observations we restrict our considerations to low values 
of $\tan\beta\lesssim 2$ that allows to obtain $|m_{\chi_1^0}|\sim |m_{\chi_2^0}|\sim M_Z/2$.
However, even for $\tan\beta\lesssim 2$ the lightest inert neutralino states can get
appreciable masses only if inert chargino mass eigenstates are light, i.e. 
$m_{\chi^{\pm}_1}\simeq 100-200\,\mbox{GeV}$. We demonstrate (see benchmark point (v) 
in Table 1) that the scenarios with only one light inert chargino mass eigenstate 
may lead to the dark matter density consistent with cosmological observations.

When $\tan\beta\lesssim 2$ the mass of the lightest CP--even Higgs boson is very 
sensitive to the choice of the coupling $\lambda(M_t)$. In particular, to satisfy LEP 
constraints $\lambda(M_t)$ must be larger than $g'_1\simeq 0.47$, where $g^{'}_1$ is 
the low energy $U(1)_{N}$ gauge coupling. If $\lambda\gtrsim g'_1$ the vacuum stability 
requires all Higgs states except the lightest one to be considerably heavier than the 
EW scale so that the qualitative pattern of the Higgs spectrum is rather similar to the 
one which arises in the PQ symmetric NMSSM \cite{Miller:2003ay}-\cite{Nevzorov:2004ge},
\cite{Miller:2005qua}. In this case the lightest Higgs state manifests itself in the 
interactions with gauge bosons and fermions as a SM--like Higgs boson.

Our benchmark scenarios indicate that in the case when 
$|m_{\chi_1^0}|\sim |m_{\chi_2^0}|\sim M_Z/2$ the SM--like Higgs boson 
decays more than 95\% of the time into $\chi^0_1$ and $\chi^0_2$ while the total 
branching ratio into SM particles varies from 2\% to 4\%. When the masses of the 
lightest and second lightest inert neutralinos are close or they form a Dirac 
(pseudo--Dirac) state (see benchmark scenarios (ii) and (v) in Table 1) then the 
decays of the lightest Higgs boson into $\chi^0_1$ and $\chi^0_2$ lead to the 
missing $E_T$ in the final state. Thus these decay channels give rise to a large 
invisible branching ratio of the SM--like Higgs boson. If the mass difference between 
the second lightest and the lightest inert neutralino is $10\,\mbox{GeV}$ or more, 
then some of the decay products of a $\chi^0_2$ that originates from a
SM-like Higgs boson decay might be observed at the LHC. In our analysis we assume that
all scalar particles, except for the lightest Higgs boson, are heavy and that the couplings 
of the inert neutralino states to quarks, leptons and their superpartners are relatively 
small. As a result the second lightest inert neutralino decays into the lightest one 
and a fermion--antifermion pair mainly via a virtual $Z$. In our numerical 
analysis we did not manage to find any benchmark scenario with 
$|m_{\chi_2^0}|-|m_{\chi_1^0}|\gtrsim 20\,\mbox{GeV}$
leading to reasonable values of $\Omega_{\mathrm{CDM}}h^2$. Hence we do 
not expect any observable jets at the LHC associated with the decay of a
$\chi^0_2$ produced through a Higgs decay. However, it might be possible 
to detect $\mu^{+} \mu^{-}$ pairs that come from the exotic decays of the 
lightest CP--even Higgs state mentioned above.

In Table 1 benchmark scenarios (i), (iii), (iv) can lead to these relatively
energetic muon pairs in the final state of the SM-like Higgs decays. Since the Higgs 
branching ratios into SM particles are rather suppressed, the decays of the
lightest CP--even Higgs state into $l^{+} l^{-} + X$ might play an essential role in
Higgs searches. 

In Table 1 we also specify the interval of variations of $\sigma_{SI}$ 
for each benchmark scenario. The lower limit on $\sigma_{SI}$ corresponds to $f^N_{Ts}=0$ 
while the upper limit implies that $f^N_{Ts}=0.36$. From Table 1 and Eq.~(\ref{higgs17}) 
it also becomes clear that $\sigma_{SI}$ decreases when $m_{h_1}$ grows.
Since in all of the benchmark scenarios presented in Tables 1 the lightest inert 
neutralino is relatively heavy $(|m_{\chi^0_{1}}|\sim M_Z/2)$, allowing for a small 
enough dark matter relic density, the coupling of $\chi_1^0$ to the lightest CP-even 
Higgs state is always large giving rise to a $\chi_1^0$--nucleon spin-independent cross 
section which is on the edge of observability of XENON100.

In addition to the exotic Higgs decays and large LSP direct detection   
cross-sections, the scenarios considered here imply that
at least two of the inert neutralino states that are predominantly
the fermion components of the inert Higgs doublet superfields and one of the 
inert chargino states should have masses below $200\,\mbox{GeV}$. Because these 
states are almost inert Higgsinos they couple rather strongly to $W$ and $Z$--bosons. 
Thus at hadron colliders the corresponding inert neutralino and chargino states 
can be produced in pairs via off-shell $W$ and $Z$--bosons. Since they are light 
their production cross sections at the LHC are not negligibly small. After being 
produced inert neutralino and chargino states sequentially decay into the LSP 
and pairs of leptons and quarks resulting in distinct signatures that can be
discovered at the LHC in the near future.

\begin{acknowledgments}
R.N. would like to thank FTPI, University of Minnesota for its hospitality and
M.~A.~Shifman, K.~A.~Olive, A.~I.~Vainshtein, A.~Mustafayev, W.~Vinci, P.~A.~Bolokhov, 
P.~Koroteev for fruitful discussions. Authors are grateful to X.~R.~Tata, M.~Muehlleitner, 
L.~Clavelli, D.~Stockinger, D.~J.~Miller, D.~G.~Sutherland, J.~P.~Kumar, D.~Marfatia,
K.~R.~Dienes, B.~D.~Thomas for valuable comments and remarks. The work of R.N. and S.P. 
was supported by the U.S. Department of Energy under Contract DE-FG02-04ER41291, 
and the work of M.S. was supported by the National Science Foundation PHY-0755262. 
S.F.K. acknowledges partial support from the STFC Rolling Grant ST/G000557/1.
J.P.H. is thankful to the STFC for providing studentship funding.
\end{acknowledgments}

\end{document}